# Dynamic Model Updating Using Particle Swarm Optimization Method

Tshilidzi Marwala (t.marwala@ee.wits.ac.za)
School of Electrical and Information Engineering
University of the Witwatersrand
P/Bag 3, Wits, 2050, South Africa

**This paper proposes the use of particle swarm optimization method (PSO) for finite element (FE) model updating. The PSO method is compared to the existing methods that use simulated annealing (SA) or genetic algorithms (GA) for FE model for model updating. The proposed method is tested on an unsymmetrical H-shaped structure. It is observed that the proposed method gives updated natural frequencies the most accurate and followed by those given by an updated model that was obtained using the GA and a full FE model. It is also observed that the proposed method gives updated mode shapes that are best correlated to the measured ones, followed by those given by an updated model that was obtained using the SA and a full FE model. Furthermore, it is observed that the PSO achieves this accuracy at a computational speed that is faster than that by the GA and a full FE model which is faster than the SA and a full FE model.**

## Introduction

Finite Element (FE) model updating entails tuning the model so that it can better reflect the measured data from the physical structure being modeled [1]. One fundamental characteristic of an FE model is that it can never be a true reflection of the physical structure but will forever be an approximation. In other words, FE updating fundamentally implies that we are identifying a better approximation of the physical structure than the original model. The aim of this paper is to introduce updating of finite element models using Response Surface Method (PSO) [2]. Thus far, the PSO method has not been used to solve the FE updating problem [1]. This new approach is compared with a method that uses simulated annealing (SA) or genetic algorithms together with a full FE model for updating. FE updating methods have been implemented using different types of optimization methods such as genetic algorithm (GA) and conjugate gradient method [3-5]. Levin and Lieven [5] proposed the use of simulated annealing (SA) and genetic algorithms (GA) for FE updating.

PSO is an approximate optimization method that looks at various design variables and their responses and identify the combination of design variables that give the best response. In this paper, the best response is defined as the one that gives the minimum distance between the measured data and the data predicted by the FE model. PSO attempts to replace implicit functions of the original design optimization problem with an approximation model, which traditionally are polynomials and are less expensive to evaluate. This makes PSO very useful to FE model updating because optimizing the FE to match measured data is a computationally expensive exercise. Furthermore, the calculation of the gradients that are essential when traditional optimization methods, such as conjugate gradient methods, are used is computationally expensive and often encounters numerical problems such as ill-conditioning. PSO tends to be immune to such problems when used for FE model updating. This is largely because PSO solves a crude approximation of the FE model rather than the full FE model which is of high dimensional order. In this paper we use the multi-layer perceptron (MLP) [6] to approximate the response equation. The PSO is particularly useful for optimizing systems that are evolving as a function of time, a situation that is prevalent in model-based fault diagnostics in the manufacturing sector. To date, PSO has been used extensively to optimize complex models and processes [7, 8].

In summary, the PSO is used because of the following reasons: (1) the relative ease of implementation that includes low computational time when compared to other methods; (2) the suitability of the approach to the manufacturing sector where model-based methods are often used to monitor structures that evolve as a function of time.

Finite element model updating has been used widely to detect damage in structures [9]. When implementing FE updating methods for damage identification, it is assumed that the FE model is a true dynamic representation of the structure. This means that changing any physical parameter of an element in the FE model is equivalent to introducing damage in that region. There are two approaches used in FE updating: direct methods and iterative methods [1]. Direct methods, which use the modal properties, are computationally efficient to implement and reproduce the measured modal data exactly. Furthermore, they do not take into account the physical parameters that are updated. Consequently, even though the FE model is able to predict measured quantities, the updated model is limited in the following ways: it may lack the connectivity of nodes - connectivity of nodes is a phenomenon that occurs naturally in finite element modeling because of the physical reality that the structure is connected; the updated matrices are populated instead of banded - the fact that structural elements are only connected to their neighbors ensures that the mass and stiffness matrices are diagonally dominated with few couplings between elements that are far apart; and there is a possible loss of symmetry of the systems matrices. Iterative procedures use changes in physical parameters to update FE models and produce models that are physically realistic. In this paper, iterative methods that use modal properties for FE updating are implemented in the context of the PSO. In this paper, FE models are updated so that the measured natural frequencies match the natural frequencies predicted by the FE model. The mode shapes are then used to cross-validate the accuracy of the model. The proposed PSO updating method is tested on an unsymmetrical H-structure.

## Mathematical Background

In this paper, modal properties, i.e. natural frequencies and mode shapes, are used as a basis of FE model updating. For this reason these parameters are described in this section. Modal properties are related to the physical properties of the structure. All elastic structures may be described in terms of their distributed mass, damping and stiffness matrices in the time domain through the following expression [10]:

$$[M]\{X''\} + [C]\{X'\} + [K]\{X\} = \{F\} \qquad (1)$$

where $[M]$, $[C]$ and $[K]$ are the mass, damping and stiffness matrices respectively, and $\{X\}$, $\{X'\}$ and $\{X''\}$ are the displacement, velocity and acceleration vectors respectively while $\{F\}$ is the applied force vector. If equation 1 is transformed into the modal domain to form an eigenvalue equation for the $i^{th}$ mode, then [10]:

$$(-\overline{\omega}_i^2[M] + j\overline{\omega}_i[C] + [K])\{\overline{\phi}\}_i = \{0\} \qquad (2)$$

where $j = \sqrt{-1}$, $\overline{\omega}_i$ is the $i^{th}$ complex eigenvalue, with its imaginary part corresponding to the natural frequency $\omega_i$, and $\{\overline{\phi}\}_i$ is the $i^{th}$ complex mode shape vector with the real part corresponding to the normalized mode shape $\{\phi\}_i$. The introduction of damage in structures changes the mass and stiffness matrices. From equation 2 it may be deduced that changes in the mass and stiffness matrices cause changes in the modal properties of the structure.

The frequency response functions (FRFs) are defined as the ratio of the Fourier transformed response to the Fourier transformed force. The FRFs may be expressed in receptance and inertance form. On the one hand, receptance expression of the FRF is defined as the ratio of the Fourier transformed displacement to the Fourier transformed force. On the other hand, inertance expression of the FRF is defined as the ratio of the Fourier transformed acceleration to the Fourier transformed force. The inertance FRF $(H)$ may be written in terms of the modal properties by using the modal summation equation as follows [10]:

$$H_{kl}(\omega) = \sum_{i=1}^{N} \frac{-\omega^2 \phi_k^i \phi_l^i}{-\omega^2 + 2\zeta_i \omega_i \omega j + \omega_i^2} \qquad (3)$$

Equation 3 is FRF due to excitation at k and measurement at l, N is the number of modes and $\zeta_i$ is the damping ratio of mode

i. Using the excitation and response of the structure and Fourier transform the FRFs may be calculated and using equation 3 and a technique called modal analysis [10], the natural frequencies and mode shapes can be indirectly calculated. The modal properties of a dynamic system depend on the mass and stiffness model of the system as indicated in equation 2. Therefore, the measured modal properties can be reproduced by the model if the mass and stiffness matrices are correctly identified. The mass and stiffness matrices can be manipulated by the manipulation of the modulus of elasticity [1]. In this paper we use the modulus of elasticity of the system as parameters that are used for finite element updating. In this paper the cost function that is minimized is as follows:

$$error = \sum_{i=1}^{N} \gamma_i \left( \frac{f_i^m - f_i^{calc}}{f_i^m} \right)^2 \qquad (4)$$

Here $m$ is for measured, $calc$ is for calculated and N is the number of modes. Here $\gamma_i$ is the weighting factor that measures the relative distance between the initial estimated natural frequency for a mode $i$ and the target frequency of the same mode. The updated model will be evaluated by comparing the natural frequencies and mode shapes before updating and after updating. The mode shapes are used for cross-validation as they were not directly linked to equation 4. Cross-validation is an important statistical technique that has proven to be useful in the area of system identification such as neural network modeling [6].

## Surface Response Method (PSO)

PSO is an evolutionary programming method that was developed by Kennedy and Eberhart [***]. This procedure was inspired by algorithms that model the "flocking behavior" seen in birds. The PSO approach has the advantages in that it is computationally efficient, simple to implement and is able to adapt to the local and global exploratory ability. When implementing the PSO the simulation is initialized with a population of random candidates, conceptualized as particles. Each particle is assigned a random velocity and is iteratively moved through the parameter space. At each step, the particle is attracted towards a region of the best fitness function by the location of the best fitness achieved so far in the population. On implementing the standard PSO, each particle is represented by two vectors: $p_i(k)$ the position and $v_i(k)$ the velocity at step $k$. Initial positions and velocities of particles are randomly generated and the subsequent positions and velocities are calculated while remembering the position of the best solution that the particle has encountered $pbest_i(k)$ and $gbest(k)$ is the best particle in the swarm. The subsequent velocity and position of a particle $i$ can be calculated using these equations:

$$v_i(k+1) = wv_i(k) + c_1 r_1(pbest_i(k) - p_i(k))$$
$$+ c_2 r_2(gbest(k) - p_i(k)) \qquad (2)$$

$$p_i(k+1) = p_i(k) + v_i(k+1) \qquad (1)$$

Here $w$ is the weight that is to control the impact of the previus velocity of the particle on the current velocity; $c_1$ and $c_2$ are parameters that emphasize individual and group experiences, respectively; and $r_1$ and $r_2$ are random numbers that determine the degree in which the simulation explore the space.

## Genetic Algorithms

GA was inspired by Darwin's theory of natural evolution. Genetic algorithm is a simulation of natural evolution where the law of the survival of the fittest is applied to a population of individuals. In this paper, this natural optimization method is used to optimize either the response surface approximation equation or the error between the FE model and the measured data. GA is implemented by generating a population and creating the new population by performing the following steps: (1) crossover; (2) mutation; (3) and reproduction. The details of these steps can be found in Holland [20] and Goldberg [21]. The crossover operator mixes genetic information in the population by cutting pairs of chromosomes at random points along their length and exchanging over the cut sections. This has a potential of joining successful operators together. Arithmetic crossover technique [21] is used in this paper. Arithmetic crossover takes two parents and performs an interpolation along the line formed by the two parents. For example if two parents $p1$ and $p2$ cross-over, then a random number $a$ which lies in the interval [0,1] is generated and the new off-springs are $p1(a-1)$ and $pa$. The mutation is a process that introduces to the population new information. In this paper non-uniform mutation [21] is used and it changes one of the parameters of the parent based on a non-uniform probability distribution. The Gaussian distribution starts with a high variance and narrows to a point distribution as the current generation approaches the maximum generation. Reproduction takes successful chromosomes and reproduces them in accordance to their fitness functions. In this paper normalized geometric selection method is used [21]. This method is a ranking selection function which is based on the normalized geometric distribution. Using this method the least fit members of the population are therefore gradually driven out of the population. The basic genetic algorithm implemented in this paper is implemented as follows:

1) Randomly create an initial population of a certain size.

2) Evaluate all of the individuals in the population using the objective function in equation 4.

3) Use the normalized geometric section method to select a new population from the old population based on the fitness of the individuals as given by the objective function.

4) Apply some genetic operators, non-uniform mutation and arithmetic crossover, to members of the population to create new solutions.

5) Repeat steps 2-6, which is one generation, until a certain fixed number of generations has been achieved

The next section describes simulated annealing which is used to update an FE model using the full FE model.

## Simulated Annealing

Simulated Annealing is a Monte Carlo method that is used to investigate the equations of state and frozen states of n degree of freedom system [22]. SA is inspired by the process of annealing where objects, such as metals, recrystalize or liquids freeze. In the annealing process the object is heated until it is molten, then it is slowly cooled down such that the metal at any given time is approximately in thermodynamic equilibrium. As the temperature of the object is lowered, the system becomes more ordered and approaches a *frozen* state at T=0. If the cooling process is conducted insufficiently or the initial temperature of the object is not sufficiently high, the system may become quenched forming defects or freezing out in metastable states. This indicates that the system is trapped in a local minimum energy state.

The process that is followed to simulate the annealing process was proposed by Metropolis *et al.* [22] and involves choosing the initial state at energy $E$ (see equation 6) and temperature $T$ and holding $T$ constant and perturbing the initial configuration computing $E$ at the new state. If $E$ at the new state is lower than $E$ at the previous state, then accept the new state, otherwise if the opposite is the case then accept this state with a probability of $exp -(dE/T)$ where $dE$ is the change in energy. This process can be mathematically represented as follows:

$$if\ E_{new} < E_{old}\ accept\ state\ E_{new}$$

$$else\ accept\ E_{new}\ with\ probability\ exp\left(\frac{E_{new} - E_{old}}{T}\right) \qquad (7)$$

This processes is repeated such that the sampling statistics for the current temperature is adequate, and then the temperature is decreased and the process is repeated until a frozen state where $T=0$ is achieved.

SA was first applied to optimization problems by Kirkpatrick, *et al.* [23]. In this paper, the current state is the current updating solution, the energy equation is the objective function in equation 4, and the ground state is the global optimum solution.

## Example: Asymmetrical H-structure

An unsymmetrical H-shaped aluminum structure shown in Figure 3 was used. This structure was also used by Marwala and Heyns [4] and by Marwala [25]. This structure had three thin cuts of 1mm that went half-way through the cross-section of the beam. These cuts were introduced in elements 3, 4 and 5. The structure with these cuts was used so that the initial FE model gives data that are far from the measured data and thereby test the proposed procedure a difficult FE updating process. The structure was suspended using elastic rubber bands. The structure was excited using an electromagnetic shaker and the response was measured using an accelerometer. The structure was divided into 12 elements. It was excited at a position indicated by double-arrows in Figure 3 and acceleration measurements were made in 15 positions indicated by single-arrows. The structure was tested freely-suspended, and a set of 15 frequency response functions were calculated. A roving accelerometer was used in the testing. The mass of the accelerometer was found to be negligible compared to the mass of the structure. The number of measured coordinates was 15, and the number of unmeasured coordinates was 24.

Thereafter, the finite element model was constructed using the Structural Dynamics Toolbox [24]. The FE model used Euler-Bernoulli beam elements. The FE model contains 12 elements. The moduli of elasticity of these elements were used as parameters to be updated. When the FE updating was implemented the moduli of elasticity was restricted to vary from 600MPa to 800MPa. The weighting factors in equation 4 were calculated for each mode as the square of the error between the measured natural frequency and the natural frequency calculated from the initial model. When the PSO, SA and GA were implemented for model updating the results shown in Table 1 were obtained.

output unit corresponding to the error in equation 4. As described before, the MLP had a hyperbolic tangent activation function in the hidden layer and linear activation function in the output layer. The PSO functional approximation via the MLP was evaluated 10 times (iterations) each time using the GA to calculate the optimum point and storing the previous optimum point in the data set for the current functional approximation. The scaled conjugate gradient method was used to train the MLP, primarily because of its computational efficiency [26]. The initial functional approximation was obtained by training the MLP on 200 training cycles and on a subsequent functional approximation, where the data had the previous optimum solution, had 5 training cycles. On using the PSO the MLP was only initialized once. The GA was implemented on a population size of 600 and 100 generations. The normalized geometric distribution was implemented with a probability of selecting the best candidate being 8%.

When SA and a full FE model was implemented for FE updating, the scale of the cooling schedule was set to 4 and the number of individual annealing runs was set to 3. When the simulation was run, the first run involved 7008 FE model calculations, in the second run 6546 FE model calculations and in the third run 5931 FE model calculations were made.

On implementing the GA and a full FE model, the same options as those that were used in the implementation of the PSO were used.

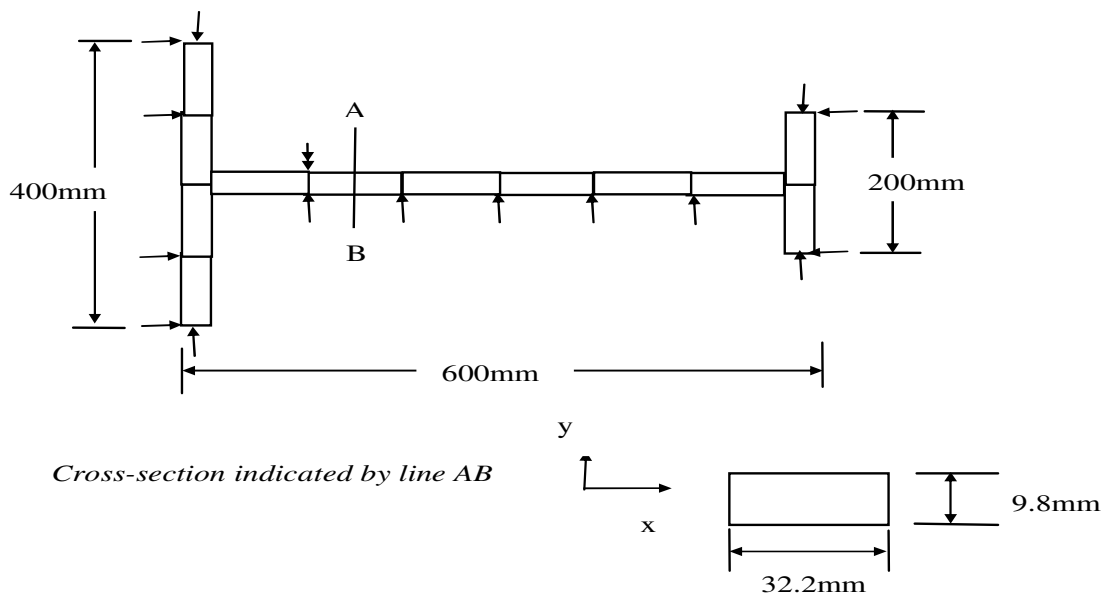

Figure 3. Irregular H-shaped structure

On implementing the proposed PSO the FE model was run 150 times to generate the data for functional approximation. The MLP implemented had 12 input variables corresponding to the 12 elements in the FE model, 8 hidden units and one

The results obtained, when the PSO, SA and GA were used for FE model updating, are shown in Table 1 are obtained. The results showing the moduli of elasticity of the initial model, PSO updated FE model, SA updated FE model and GA updated FE model are shown in Figure 4. Table 1 shows the measured natural frequencies, initial natural frequencies and natural frequencies obtained by the PSO, SA and GA updated FE models. The error between the first measured natural frequency and that from the initial FE model, which was obtained when the modulus of elasticity of 700 MPa was assumed, is 4.0%. When the PSO was used for FE updating, this error is reduced to 0.5% while using SA it is worsened to 4.5% and using the GA approach it is improved to 2.8%. The error between the second measured natural frequency and that from the initial model was 8.4%. When the PSO was used, this error was improved to 1.8% while using SA it was improved to 0.9% and using the GA it was improved 1.8%. The error between the third natural frequencies between the measured data and the initial model obtained when the standard modulus of elasticity was assumed for the initial model is 9.60%. When the PSO was used, this error was reduced to 0.38% while using SA it was reduced to 0.38% and using the GA and a full FE model was reduced to 4.65%. The error between the fourth measured natural frequency and that from the initial model was 3.70%. When the PSO was used for FE updating, this error was reduced to 2.91% while using the SA increased it to 4.81% and using the GA and a full FE model was reduced to 2.44%. The error between the fourth measured natural frequency and that from the initial model was 1.64%. When the PSO was used, this error was increased to 4.11% while using SA it increased to 6.70% and using the GA and a full FE model was increased to 4.49%. Overall, the PSO gave the best results with an average error, calculated over all the five natural frequencies, of 1.95%, followed by the GA with an average error of 3.23% and then finally the SA with an average error of 3.49%. All the three methods improved when compared to the average error between the initial FE model and the measured data of 5.51%.

Table 1. Results showing measured frequencies, the initial frequencies and the frequencies obtained when the FE model is updated using the PSO, SA and GA

| Measured Freq (Hz) | Initial Freq (Hz) | Frequencies from PSO Updated Model (Hz) | Frequencies from SA Updated Model (Hz) | Frequencies from GA Updated Model (Hz) |
|---|---|---|---|---|
| 53.9 | 56.2 | 53.9 | 53.9 | 54.0 |
| 117.3 | 127.1 | 117.8 | 118.4 | 122.0 |
| 208.4 | 228.4 | 208.5 | 209.9 | 213.2 |
| 254.0 | 263.4 | 253.8 | 253.4 | 254.5 |
| 445.1 | 452.4 | 438.5 | 437.0 | 434.9 |

The updated models implemented in this paper were cross-validated on the mode shapes they predicted. To make this assessment possible the coordinate modal assurance criterion (COMAC) [27] was used. The COMAC is a measure of the correlation between two sets of mode shapes of the same dimension. The COMAC reduces the dimension of the mode shapes to be equal to the number of degrees of freedom measured. When two sets of mode shapes are perfectly correlated then the COMAC for all degrees of freedom is 1. Otherwise, when perfectly uncorrelated, the COMAC for all degrees of freedom is 0. In this paper, the mean of the COMAC vector is used to compare the FE model predicted mode shapes before and after updating by the three methods. The average COMAC calculated between the mode shapes from an initial FE model and the measured mode shapes was 0.79. When the average COMAC was calculated between measured data and data obtained from the FE models that were updated using the PSO, SA and GA, the resulting averages of calculated from the respective COMAC vectors were 0.79, 0.9207 and 0.8650, respectively. This indicates that the PSO gives the best results for the current application, followed by the SA and then the GA.

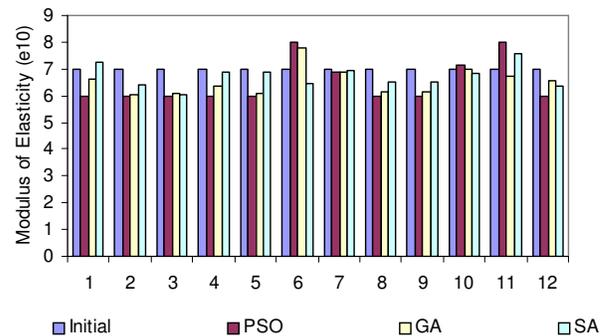

Figure 4. Graph showing the initial moduli of elasticity and the moduli of elasticity obtained when the FE model is updated using the PSO, GA and SA. Here e10 indicates 10 to the power 10.

The computational time taken to run the complete PSO method was 19 CPU minutes, while the SA and a full FE model took 21 CPU minutes to run and the GA and a full FE model took 17 CPU minutes. The PSO was found to be faster than the GA which was in turn much faster than the SA.

### Conclusion

In this paper, PSO is proposed for FE model updating. The proposed PSO was implemented within the framework of the MLP for functional approximation and GA for optimization of the MLP function. This procedure was compared to the GA and SA that were implemented independently in conjunction with the FE models. When these techniques were tested on the unsymmetrical H-shaped structure, it was observed that the PSO performed the best from both the accuracy of prediction of the modal properties and the computational load, followed by the GA which was in turn better than the SA. However, on the correlation of mode shapes it was observed that the PSO gave the best correlation between the measured mode shapes and the mode shapes predicted by updated FE models followed by the SA and then followed by the GA.

### Acknowledgment

The author would like to thank Stefan Heyns, the now National Research Foundation as well as the AECI, Ltd for their assistance in this work.

## References

[1]Friswell, M.I., and Mottershead, J.E., *Finite element model updating in structural dynamics,* Kluwer Academic Publishers Group, Norwell, MA, 1995, pp. 1-286.

[2]Montgomery, D.C., *Design and analysis of experiments,* 4th Edition, John Wiley and Sons, NY, 1995, Chapter 14.

[3]Marwala, T., "Finite element model updating using wavelet data and genetic algorithm" *Journal of Aircraft,* 39(4), 2002, pp. 709-711.

[4]Marwala, T., and Heyns, P.S., "A multiple criterion method for detecting damage on structures" *AIAA Journal,* 195(2), 1998, pp. 1494-1501.

[5]Levin, R.I. and Lieven, N.A.J, "Dynamic finite element model updating using simulated annealing and genetic algorithms" *Mechanical Systems and Signal Processing,* 12(1), pp. 91-120.

[6]Bishop, C.M., *Neural Networks for Pattern Recognition.* Oxford: Clarendon, 1996.

[7]Lee, S.H., Kim, H.Y., and Oh, S.I., "Cylindrical tube optimization using response surface method based on stochastic process" *Journal of Materials Processing Technology,* 130-131(20), 2002, pp. 490-496.

[8]Edwards, I.M., and Jutan, A., "Optimization and control using response surface methods" *Computers & Chemical Engineering,* 21(4), 1997, pp. 441-453.

[9]Doebling, S.W., Farrar, C.R., Prime, M.B., and Shevitz, D.W., "Damage identification and health monitoring of structural and mechanical systems from changes in their vibration characteristics: a literature review" *Los Alamos National Laboratory Report LA-13070-MS,* 1996.

[10]Ewins, D.J., *Modal testing: theory and practice,* Research Studies Press, Letchworth, U.K, 1995.

[11]Sacks, J., Welch, W.J., Mitchell, T.J., and Wynn, H.P., "Design and analysis of computer experiments" *Statistical Science,* 4(4), 1989, pp. 409-435.

[12]Varaajan, S., Chen, W., and Pelka, C.J., "Robust concept exploration of propulsion systems with enhanced model approximation" *Engineering Optimization,* 32(3), 2000, pp. 309-334.

[13]Giunta, A. A., and Watson, L. T., "A Comparison of Approximation Modeling Techniques: Polynomial Versus Interpolating Models," *AIAA-98-4758, American Institute of Aeronautics and Astronautics, Inc.,* 1998, pp.392-401.

[14]Koch, P. N., Simpson, T. W., Allen, J. K., and Mistree, F., 1999, "Statistical Approximations for Multidisciplinary Design Optimization: The Problem of Size," *Journal of Aircraft,* 36(1), 1999, pp. 275-286.

[15]Jin, R., Chen, W., and Simpson, T., 2000, "Comparative Studies of Metamodeling Techniques under Multiple Modeling Criteria," *8th AIAA/NASA/USAF/ISSMO Symposium on Multidisciplinary Analysis and Optimization,* Long Beach, CA, September 6-8, 2000.

[16]Lin, Y., Krishnapur, K., Allen, J. K., and Mistree, F., 2000, "Robust Concept Exploration in Engineering Design: Metamodeling Techniques and Goal Formulations," *Proceedings of the 2000 ASME Design Engineering Technical Conferences,* DETC2000/DAC-14283, September 10-14, 2000, Baltimore, Maryland.

[17]Wang, G.G., 2003, "Adaptive response surface method using inherited Latin hypercube design points," *Transactions of the ASME, Journal of Mechanical Design,*125, pp. 210-220.

[18]Simpson, T. W., Peplinski, J. D., Koch, P. N., and Allen, J. K., "Metamodels for Computer-based Engineering Design: Survey and Recommendations," *Engineering with Computers,* 17, 2001, pp. 129-150.

[19]Fletcher, R., *Practical Methods of Optimization.* 2nd edition, New York: Wiley, 1987.

[20]Holland, J, *Adaptation in Natural and Artificial Systems,* Ann Arbor: University of Michigan Press, 1975.

[21]Goldberg, D.E., *Genetic algorithms in search, optimization and machine learning,* Addison-Wesley, Reading, MA, 1989.

[22]Metropolis, N, Rosenbluth, A.W., Rosenbluth, M.N., Teller, A.H., and Teller, E., "Equations of state calculations by fast computing machines," *Journal of Chemical Physics,* 21, 1953, pp. 1087-1092.

[23]Kirkpatrick, S., Gelatt, C.D., and Vecchi, M.P., "Optimization by simulated annealing," *Science,* 220, 1983, pp. 671-680.

[24]Balmès, E., *Structural Dynamics Toolbox User's Manual* Version 2.1, Scientific Software Group, Sèvres, France, 1997.

[25]Marwala, T, *A multiple criterion updating method for damage detection on structures.* University of Pretoria Masters Thesis, 1997.

[26]Møller, M. "A scaled conjugate gradient algorithm for fast supervised learning" *Neural Networks,* vol. 6, 1993, pp. 525-533.

[27]Lieven, N.A.J. and Ewins, D.J., "Spatial correlation of mode shapes, the co-ordinate modal assurance criterion" *Proceedings of the 6th International Modal Analysis Conference,* 1988, pp. 690-695.